\documentclass[11pt]{scrartcl}
\usepackage[utf8]{inputenc}
\usepackage{comment}
\usepackage{amsthm}
\usepackage{amsmath}
\usepackage{amssymb}
\usepackage{array}
\usepackage{tabularx}
\usepackage{graphicx}
\usepackage{adjustbox}
\usepackage{booktabs}
\usepackage{caption}
\newtheorem{remark}{Remark}
\usepackage[ruled,vlined]{algorithm2e}
\usepackage[table,x11names]{xcolor}

\usepackage{subcaption}
\usepackage{authblk}
\newcommand\lword[1]{\leavevmode\nobreak\hskip0pt plus\linewidth\penalty50\hskip0pt plus-\linewidth\nobreak\textbf{#1}}

\usepackage[colorlinks = true,
            linkcolor = blue,
            urlcolor  = blue,
            citecolor = blue,
            anchorcolor = blue]{hyperref}
\usepackage[authoryear,round,numbers]{natbib}
\newcommand*\samethanks[1][\value{footnote}]{\footnotemark[#1]}

\title{\vspace*{-1\baselineskip}Multi-Period Portfolio Optimization using Model Predictive Control with Mean-Variance and Risk Parity Frameworks}
\author[1]{Xiaoyue Li\thanks{These two authors contributed equally}\thanks{\url{xiaoyuel@princeton.edu}}}
\author[1]{A. Sinem Uysal\samethanks[1]\thanks{\url{auysal@princeton.edu}}}
\author[1]{John M. Mulvey\thanks{The Bendheim Center for Finance, Center for Statistics and Machine Learning, Princeton University}}

\affil[1]{Department of Operations Research and Financial Engineering, Princeton University}
\date{\today}
\addtokomafont{author}{\vspace*{-1em}}
\addtokomafont{date}{\vspace*{-1em}}
\begin{document}
\maketitle




\begin{abstract}
We employ model predictive control for a multi-period portfolio optimization problem. In addition to the mean-variance objective, we construct a portfolio whose allocation is given by model predictive control with a risk-parity objective, and provide a successive convex program algorithm that provides 30 times faster and robust solutions in the experiments.
Computational results on the multi-asset universe show that multi-period models perform better than their single period counterparts in out-of-sample period, 2006-2020. 
The out-of-sample risk-adjusted performance of both mean-variance and risk-parity formulations beat the fix-mix benchmark, and achieve Sharpe ratio of 0.64 and 0.97, respectively.  

\end{abstract}
\newpage
\section{Motivation/Introduction}
Portfolio optimization is one of the central problems in financial engineering. Popular allocation methods include mean-variance optimization (\cite{Markowitz1952}), fixed-mix strategies motivated by \cite{Merton1969}, risk parity and risk budgeting (\cite{Maillard2010}, \cite{Bruder2012}, \cite{Chaves2011}), etc. Fixed-mix strategies such as 60/40 provides a rule-of-thumb for average investors who seek for a steady growth of the portfolio over a long horizon with modest downside protections. Mean-variance optimization takes advantage of information by considering the estimation and forecasts of the performance of underlying assets. It successfully catches the uptrend when one has promising mechanism for predicting future returns. On the other hand, empirical results suggest that the estimation of asset returns is difficult and inaccurate compared to the estimation of covariance matrices. To alleviate the harm of inferior return predictions, some investors look for strategies that are robust under worst-case scenarios (\cite{GULPINAR2007981}), and some others seek a stable investment strategy that distributes risks evenly to each of the assets, namely risk parity.

Mean-variance and risk parity allocation strategies with rolling single-period optimization have proven to be successful and elegant methods. On the other hand, single-period models fail to fully address important issues in portfolio management, including transaction costs, change of market dynamics, intermediate cash flows, and goal-based risk measures etc. \cite{Topaloglou2007} compare a dynamic stochastic programming model for portfolio management with single-stage and two-stage setups, and find that the two-stage model provides a dominating return-CVaR efficient frontier. In this paper, we introduce a portfolio allocation framework based on multi-period optimization problem with selection of allocation model.

It is non-trivial to solve multi-period optimization problems. With traditional numerical methods, the running time grows exponentially as a function of problem size. Even a modest number of time periods, say above 5-10, and branching processes exceeds the capability of most modern computers without specialized algorithms. To deal with the curse of dimensionality, we adopt a model predictive control (MPC) framework for portfolio optimization (\cite{Boyd2017}), and compare the performance of MPC based on mean-variance optimization versus that based on risk parity. The market environment is described by a  hidden Markov model (\cite{Nystrup2019}) which provides necessary asset parameter forecasts for multi-period optimization models. Our results show that the models have different strengths and their performance depends upon the underlying market environment, but both succeed in beating the fix-mix benchmark. 
The investment framework we propose is as follows:
\begin{enumerate}
    \item Employ the hidden Markov model on the historical returns to estimate the expected returns and covariance matrices for future periods;
    \item Solve two multi-period portfolio allocation problems based on MPC with mean-variance objective and MPC with risk-parity objective. 
\end{enumerate}
The contribution of this paper is two-fold. First, we extend model predictive control approach from \cite{Boyd2017} to solve a multi-period risk-parity portfolio with transaction control. The resulting portfolio enjoys a high Sharpe ratio and an evenly distributed risk contribution compared to benchmark strategies and the mean-variance approach. We also illustrate the asset allocations of both mean-variance and risk-parity strategies under specific a time period to show the relative strength of each objective. Second, a successive convex algorithm is derived to solve the MPC problem with a risk-parity objective. The convex formulation of risk-parity objective shortens the running time and stabilizes the numerical solution. We find that empirically, the algorithm converges within a few steps and offers desired global solution.

Section \ref{HMM} introduces the hidden Markov model and our implementation to address the market environment for the next time step. Section \ref{MPO} introduces the multi-period portfolio optimization problem,  and Section \ref{MPC} explains model predictive control approach with the mean-variance and risk parity formulations. Section \ref{Empirical} presents computational results, and Section \ref{conclusion} concludes.

\section{Hidden Markov Model} \label{HMM}
We address the market returns by a hidden Markov model (HMM) (\cite{Baum1970}). In this section, we discuss the structure of HMM and the merit of it. Popular pricing and return models include geometric Brownian motion, mean-reverting models such as 
Ornstein–Uhlenbeck process, regime-based model such as HMM, etc. Elegant as geometric Brownian motion, market returns are often asymmetric and have a heavier left tail than a normal distribution, suggesting that a HMM could provide a more realistic description of the return distribution. \cite{Dias2014} discuss the regime classification for 21 stock markets, and find their extended HMM provides meaningful classification for general markets. In fact, the usefulness of HMM has been proven by successfully capturing the stock price change during the 2008 financial crisis. Many papers implement HMM to utilize regime-switching dynamics in investment decision making framework, including examples from \cite{Bae2014}, \cite{Reus2016} and \cite{Nystrup2019}. The recent paper from \cite{UysalMulvey} show benefits of regime-switching models in risk parity portfolios.

\subsection{Hidden Markov Model}
A hidden Markov model is a pair of series $\{(X_t),(Y_t)\}$ such that 
\begin{itemize}
    \item $\{(X_t)\}$ is a Markov process, whose states are not directly observable; and 
    \item $\{(Y_t)\}$ is a series of $X_t$-measurable variables that can be observed.
\end{itemize}
For simplicity and interpretability, we will model the market returns with two states: \textit{normal regime} and \textit{contraction regime}. In each of the regimes, the returns follow a multivariate Gaussian distribution whose parameters depends on the underlying regime. In the context of HMM, $\{(X_t)\}$ is the series of market regime which the investors cannot perceive directly; $\{(Y_t)\}$ is the return series that investors may observe and by which they infer the likelihood of the current regime.

During normal regimes, the market usually experiences positive expected returns and low (co)variances, whereas in contraction regimes, the market faces low expected returns and high (co)variances. The transitions between these two regimes are described by a probabilistic matrix 
$\begin{bmatrix}
p_{nn} & 1-p_{nn}\\
1-p_{cc} & p_{cc}
\end{bmatrix}$,
where $p_{nn}$ is the probability that the next period stays in normal regime given the current period is normal, and $p_{cc}$ is the probability that the next period stays in contraction regime if the current period is contraction.

The forecasting of future returns and covariance matrices are based on fitted HMM parameters. Given that the returns follow $\mathcal{N}(\mu_n,\Sigma_n)$ under normal regime and $\mathcal{N}(\mu_c,\Sigma_c)$ under contraction regime, and that the probability of the market being normal at time $t$ is $q_t$, then
\begin{itemize}
    \item the probability that the market is under normal regime at time $t+1$ is $\hat{q}_{t+1} = q_t p_{nn} + (1-q_t)(1-p_{cc})$;
    \item the forecasting of expected return vector at time $t+1$ is $\hat{\mu}_{t+1} = \hat{q}_{t+1}\mu_n + (1-\hat{q}_{t+1})\mu_c$; 
    \item the forecasting of covariance matrix of returns at time $t+1$ is $\hat{\Sigma}_{t+1} = (\hat{q}_{t+1}\Sigma_n + (1-\hat{q}_{t+1})\Sigma_c) + (\hat{q}_{t+1}(\mu_n-\hat{\mu}_{t+1})^2 + (1-\hat{q}_{t+1})(\mu_c-\hat{\mu}_{t+1})^2)$.
\end{itemize}
The parameters of time $t+2, t+3, ...$ can be estimated with $\hat{q}_{t+1}, \hat{\mu}_{t+1}$ and $\hat{\Sigma}_{t+1}$, and iteratively.

\subsection{Our Implementation of Hidden Markov Model}
Importantly, in our empirical work, we employ approaches that are evaluated in an out of sample test period, including estimating the parameters of the hidden Markov model (HMM). We estimate the parameters of the HMM by the expectation-maximization algorithm based on market returns over the past 2000 days. The rolling window is chosen to be 2000 days so that (i) the estimation is made on the most recent market performance, and (ii) the data is likely to include a full market cycle. 
To illustrate the effectiveness of HMM, we compare the performance of portfolios based on the parameters from HMM versus those based on the naive historical estimates.
For stability of the estimation, we fit the regimes with two driving market variables - US equities and US treasury returns, and use the regime labels to estimate the mean and covariance matrix under each regime for all asset categories. Here, a regime is label in the growth period if the HMM probability is equal to .5 or greater, whereas the regime is taken as contraction if the probability is less than .5 for the next period. 

\section{Multi-period Portfolio Optimization}\label{MPO}
Effective as single-period portfolio optimization models, they fail to adequately address critical concerns in portfolio management, including transaction costs and taxes, the change of asset return dynamics, and short-term versus long-term benefit trade-off. Here, we provide a generic model for multi-period portfolio optimization problem as follows:
\begin{align}
\begin{split}
    \underset{\pi_0,\pi_1,...,\pi_{T-1}\in\mathbb{R}^n}{\text{Maximize }}& \hspace{1cm} Utility[Z_1,Z_2,...]\\
    \text{subject to} &\hspace{1cm} \mathbf{1}^T\pi_t = 1 \hspace{0.5cm} \forall t=0,...,T-1\\
    &\hspace{1cm} W_{t}^{\rightarrow} = W_{t} (\pi_t^T(1+r_t)) \hspace{0.5cm} \forall t=0,...,T-1\\
    &\hspace{1cm} \pi_{t}^\rightarrow = \frac{\pi_t\odot(1+r_t)}{\pi_t^T(1+r_t)} \hspace{0.5cm} \forall t=0,...,T-1\\
    &\hspace{1cm} W_{t+1} = W_{t}^{\rightarrow} - C(W_{t}^{\rightarrow}; \pi_{t}^\rightarrow, \pi_{t+1}) \hspace{0.5cm} \forall t=0,...,T-1\\
\end{split}
\end{align}
where $T$ is the investment horizon, $\pi_0,\pi_1,...,\pi_{T-1}\in\mathbb{R}^n$ are the allocation at the beginning of each period, $\mathbf{1}\in\mathbb{R}^n$ is the vector with all ones, $W_t$ is the dollar value of wealth at the beginning of period $t$, $r_t\in\mathbb{R}^n$ is the vector of returns in period $t$, $\odot$ is the element-wise multiplication operator, and $C(W; \pi^\rightarrow,\pi)$ is the dollar value of transaction and market impact costs when the allocation is rebalanced to $\pi^\rightarrow$ from $\pi$ with current wealth being $W$. All quantities with ${ }^\rightarrow$ correspond to the quantity at the end of the period. Initial wealth $W_0$ and the transaction function $C(\cdot;\cdot,\cdot)$ are given. The distribution of asset returns $r_t$ depends on specific circumstances. The arguments of utility function $Z_1, Z_2,...$ can be, but are not limited to, expected terminal wealth $\mathbb{E}[W_T]$, volatility of terminal wealth $std[W_T]$, and so on.

Multi-period models provide superior capabilities over single-period models, as it takes into consideration of future events when planning for the current period. However, multi-period optimization suffers from the curse of dimensionality, the phenomenon that running time grows exponentially as the complexity of problem increases. In the following section, we will introduce a truncated multi-period model that successfully avoids the curse of dimensionality while taking advantage of multi-period forecasting.

\section{Model Predictive Control}\label{MPC}
\cite{Boyd2017} propose a stochastic control formulation of the  multi-period portfolio optimization problem and employ model predictive control (MPC) to find an investment policy in finite horizon. In the  MPC approach, all unknown parameters (asset returns, covariance matrix etc.) are replaced by their forecasted values over a planning horizon. This turns the stochastic control problem into a multi-period deterministic optimization problem. Although the MPC approach yields a sub-optimal investment policy, it is computationally efficient, and incorporates  time-varying estimates of the parameters. Furthermore, \cite{Boyd2014} show that MPC policies' performances are close to optimal in the simulation studies. \cite{Boyd2017} and \cite{Nystrup2019} formulate the multi-period portfolio problem with a mean-variance objective term, including transaction and holding constraints. Recently \cite{Oprisor2020} implement the similar multi-period portfolio with Black-Litterman approach in parameter estimates. 

We consider two variations of the multi-period portfolio problem with MPC approach: 1) mean-variance, and 2) risk-parity.
The mean-variance framework has been the traditional approach to decide portfolio allocations on the basis of return-risk trade-off (\cite{Markowitz1952}). The mean-variance coefficient provides an intuitive choice of return-risk balance, and investors may choose the coefficient according to their risk appetite. \cite{Platanakis2020} show that mean-variance portfolios are superior to 1/n portfolios the in asset allocation problem. However, it faces practical drawbacks (\cite{Kolm2014}), some of which are, sensitivity to estimated input parameters and  concentration of portfolio risk. Among input parameters, expected return estimates cause large changes in asset allocations and they are found to be hard to estimate. On the other hand, \cite{Pieter2021} provide an extensive study of robustness of mean-variance portfolios under different modeling choices, and they find that dynamic mean-variance portfolios are more robust to model misspecifications, especially in the multi-period setting.

Risk management has become an important part in portfolio analysis, especially since the 2008 financial crisis. Investors move towards more risk-based investment strategies due to their historical low drawdown performance. Risk budgeting portfolio optimization is a popular risk-based asset allocation  technique (\cite{Bruder2012}). Here, the risk budgets are assigned to each assets' risk contribution, and equalizing all risk budgets in the portfolio is known as risk parity strategy (\cite{Maillard2010}). Unlike mean-variance, risk parity strategy provides a balanced risk concentration in the portfolio. Furthermore, risk parity portfolios don't require expected asset return estimates as input, and it's shown that they are robust to estimation errors.  By its nature, the risk parity portfolios tilt towards primarily in low risk assets (historically fixed income products), and provide low but stable returns over the investment horizon. Although risk parity portfolios can be conservative from an investor perspective, that can be enhanced with leverage and target return constraints. There are conflicting views of the risk parity performance in high interest rate environment due to fixed-income heavy asset allocation. However, \cite{UysalMulvey} show that regime-switching models can help to improve the risk parity performance in the historical high interest rate period of the late 1970s. Some researchers (\cite{Chaves2011}) show that the performance of the risk parity is dependent on the selected asset universe. On the computational aspect, finding risk parity portfolio allocations is complicated due to the non-convex nature of the optimization problem. Even though \cite{Maillard2010} provide a convex formulation in the case of long-only budget constraints, it doesn't generalize into the other portfolio choices. 

To demonstrate the performance and compare relative strengths, we implement multi-period formulations based on both frameworks.
\subsection{Model Predictive Control with Mean-Variance}
We introduce the mean-variance term and transaction cost with  $\ell_1$ penalty in the objective function with long-only budget constraints as follow:
\begin{equation}
    \begin{aligned}
        \underset{\pi_{t+1},\dots,\pi_{t+H} \in \mathbb{R}^n}{Maximize} \quad & \sum_{\tau=t+1}^{\tau = t+H} \hat{r}_{\tau|t}^T\pi_\tau - \gamma^{risk} (\pi_\tau^T \hat{\Sigma}_{\tau|t}\pi_\tau)-\gamma^{trade}\|\pi_\tau-\pi_{\tau-1}\|_1\\
        & \pi_\tau \geq 0 \quad \forall \tau = t+1,\dots, t+H\\
        & \mathbf{1}^T\pi_\tau = 1 \quad \forall \tau = t+1,\dots, t+H\\
    \end{aligned}
    \label{mpo}
\end{equation}
where $\gamma^{risk}$ is the risk-aversion parameter and $\gamma^{trade}$ is the penalty for transactions. When transaction costs incur, the transaction penalty helps to avoid unnecessary turnovers associated with weak signals, which is intuitively consistent with \cite{Pedersen2013} who suggest a gradual move toward the target portfolio under the existence of transaction costs. $\hat{r}_{\tau|t}$ and $\hat{\Sigma}_{\tau|t}$ are return and covariance matrix forecasts by HMM at time $t$ for periods $\tau = t+1,\dots,t+H$. Note that the actual transaction cost is calculated with the difference between the beginning-of-period allocation and the end-of-period allocation of previous period. Here, in the formulation, for calculation simplicity, we replace the end-of-period allocation of previous period with the beginning-of-period allocation of previous period. When each time period is short, it is reasonable to assume that the end-of-period allocation is close to that at the beginning of period. Therefore, $-\gamma^{trade}\|\pi_\tau-\pi_{\tau-1}\|_1$ provides a meaningful control of portfolio turnovers.\\
\\
At each period $t$, we solve the multi-period Problem \ref{mpo} over the planning horizon $H$, which produces allocation vectors $\pi_{t+1}, \dots, \pi_{t+H}$. We execute the first trade ($\pi_{t+1}$) and move on to the next period, and solve the problem until the end of investment horizon $T$. Note that at period $t$, $\pi_t$ is not a decision variable, it is an input which denotes the current portfolio allocation. Notice that Problem \ref{mpo} has a convex objective function with linear constraints, and it is efficiently solved by publicly available convex program solvers.    
\subsection{Model Predictive Control with Risk Parity}
Following the same formulation structure in Problem \ref{mpo}, we implement risk parity condition with volatility risk measure. Volatility, a common risk measure in practice, $\mathcal{R}(\pi) = \sqrt{\pi^T\Sigma \pi}$ is a homogeneous function of degree one which verifies Euler decomposition. Under this risk measure, the marginal risk contribution of asset $i$ is $\partial_i \mathcal{R}= (\Sigma \pi)_i/\sqrt{\pi^T\Sigma \pi}$, then risk contribution of the asset $i$ will be $ RC_i(\pi) = \pi_i\partial_i \mathcal{R}= \pi_i .\frac{(\Sigma\pi)_i}{\sqrt{\pi^T\Sigma \pi}}$. To enforce the risk parity condition in the asset allocation decision, where all risk contributions are equal, we introduce the following sum of squares term into the objective function: 
\begin{equation}
    \begin{aligned}
        \underset{\pi_{t+1},\dots,\pi_{t+H} \in \mathbb{R}^n}{Minimize} \quad & \sum_{\tau=t+1}^{\tau = t+H} \sum_{i=1}^n \left(\frac{\pi_{\tau,i}(\hat{\Sigma}_{\tau|t} \pi_\tau)_i}{\pi_\tau\hat{\Sigma }_{\tau|t}\pi_\tau} - b_i \right)^2 +\gamma^{trade}\|\pi_\tau-\pi_{\tau-1}\|_1\\
        & \pi_\tau \geq 0 \quad \forall \tau = t+1,\dots, t+H\\
        & \mathbf{1}^T\pi_\tau = 1 \quad \forall \tau = t+1,\dots, t+H\\
    \end{aligned}
    \label{prob:mpo_rp_leastsq}
\end{equation}
where $b_i \in [0,1]$ denotes risk budget for asset $i$ and it is equal to $1/n$ for risk parity portfolio. Likewise to Problem \ref{mpo}, the last part penalizes transactions with an $l_1$ term. Unlike mean-variance problem, the MPC approach in the multi-period risk parity portfolio formulation leads to a non-convex problem due to the risk parity term. 

\subsubsection{Convex formulation}
Problem \ref{prob:mpo_rp_leastsq} is non-convex due to the risk parity term. In practice, non-convexity does not only increase the running time, but also impairs the stability of results when multiple local solutions exist. In order to solve the model predictive control with risk parity efficiently and effectively, we propose a successive convex optimization algorithm for solving Problem \ref{prob:mpo_rp_leastsq}, based on the techniques introduced by \cite{Feng2015}. Let $g_{\tau,i}(\pi_{\tau,i})$ denote the deviation from desired risk budget of asset $i$ in period $\tau$ with allocation vector $\pi_{\tau,i}$, i.e., $g_{\tau,i}(\pi_{\tau,i}) = \frac{\pi_{\tau,i}(\hat{\Sigma}_{\tau|t} \pi_\tau)_i}{\pi_\tau\hat{\Sigma }_{\tau|t}\pi_\tau} - b_i$. Then we can write Problem \ref{prob:mpo_rp_leastsq} as 
\begin{equation}
    \begin{aligned}
        \underset{\pi_{t+1},\dots,\pi_{t+H} \in \mathbb{R}^n}{Minimize} \quad & \sum_{\tau=t+1}^{\tau = t+H} \sum_{i=1}^n [g_{\tau,i}(\pi_{\tau,i})]^2 +\gamma^{trade}\|\pi_\tau-\pi_{\tau-1}\|_1\\
        & \pi_\tau \geq 0 \quad \forall \tau = t+1,\dots, t+H\\
        & \mathbf{1}^T\pi_\tau = 1 \quad \forall \tau = t+1,\dots, t+H\\
    \end{aligned}
\end{equation} 

We start the successive procedure with an initial solution $\pi^0 = [\pi^0_{t+1} ,..., \pi^0_{t+H}]$. At the $k$-th iteration, a linear expansion of $g_{\tau,i}$ around $\pi_\tau^k$ gives $g_{\tau,i}(\pi_\tau)\approx g_{\tau,i}(\pi_\tau^k) + (\nabla g_{\tau,i}(\pi_\tau^k))^T(\pi_\tau-\pi_\tau^k)$, where $\nabla g_{\tau,i}$ is the gradient. We aim to solve 
\begin{equation}
    \begin{aligned}
        \underset{\pi_{t+1},\dots,\pi_{t+H} \in \mathbb{R}^n}{Minimize} \quad & \sum_{\tau=t+1}^{\tau = t+H} \left(\sum_{i=1}^n [g_{\tau,i}(\pi_\tau^k) + (\nabla g_{\tau,i}(\pi_\tau^k))^T(\pi_\tau-\pi_\tau^k)]^2 + \frac{\delta_\tau}{2}\|\pi_\tau-\pi_\tau^k\|_2^2 \right) +\gamma^{trade}\|\pi_\tau-\pi_{\tau-1}\|_1\\
        & \pi_\tau \geq 0 \quad \forall \tau = t+1,\dots, t+H\\
        & \mathbf{1}^T\pi_\tau = 1 \quad \forall \tau = t+1,\dots, t+H\\
    \end{aligned}
\end{equation}
where $\delta_\tau>0$ is a regularization term for convergence purpose. Expanding the sum of squares, one finds that the first part of the objective function can be written as 
\begin{equation}
\begin{aligned}
     &\sum_{\tau=t+1}^{\tau = t+H} \left(\sum_{i=1}^n [g_{\tau,i}(\pi_\tau^k) + (\nabla g_{\tau,i}(\pi_\tau^k))^T(\pi_\tau-\pi_\tau^k)]^2 + \frac{\delta_\tau}{2}\|\pi_\tau-\pi_\tau^k\|_2^2\right)\\
    = & \sum_{\tau=t+1}^{\tau = t+H}\left(\|A^k_\tau(\pi_\tau-\pi_\tau^k) + g_{\tau}(\pi_\tau^k)\|_2^2 + \frac{\delta_\tau}{2}\|\pi_\tau-\pi_\tau^k\|_2^2\right)\\
     &\text{where } A^k_\tau = [\nabla g_{\tau,1}(\pi_\tau^k), ..., \nabla g_{\tau,n}(\pi_\tau^k)]^T \text{ and } g_{\tau}(\pi_\tau^k) = [g_{\tau,1}(\pi_\tau^k), ..., g_{\tau,n}(\pi_\tau^k)]^T\\
    = & \sum_{\tau=t+1}^{\tau = t+H} \left(\|A_\tau^k\pi_\tau\|_2^2 + \frac{\delta_\tau}{2}\|\pi_\tau\|_2^2 + 2\pi_\tau^T(A_\tau^k)^T g(\pi_\tau^k) - 2\pi_\tau^T(A_\tau^k)^T A_\tau^k\pi_\tau^k - \delta_\tau\pi_\tau^T\pi_\tau^k \right) + constant\\
    = & \sum_{\tau=t+1}^{\tau = t+H} \left(\frac{1}{2}\pi_\tau^T Q_\tau^k\pi_\tau + \pi_\tau^T q_\tau^k\right) + constant\\
    & \text{where } Q_\tau^k = 2(A_\tau^k)^T A_\tau^k +\delta_\tau I \text{ and } q_\tau^k = 2(A_\tau^k)^T g_\tau(\pi_\tau^k) - Q_\tau^k\pi_\tau^k
\end{aligned}
\end{equation}

Note that by the form of $Q_\tau^k$, all of them are positive definite when $\delta_\tau>0$, and therefore the objective is rearranged to a convex quadratic function. In particular, at the $k$-th iteration, the problem becomes 
\begin{equation}
    \begin{aligned}
        \underset{\pi_{t+1},\dots,\pi_{t+H} \in \mathbb{R}^n}{Minimize} \quad & \sum_{\tau=t+1}^{\tau = t+H} \left(\frac{1}{2}\pi_\tau^T Q_\tau^k\pi_\tau + \pi_\tau^T q_\tau^k\right) +\gamma^{trade}\|\pi_\tau-\pi_{\tau-1}\|_1\\
        & \pi_\tau \geq 0 \quad \forall \tau = t+1,\dots, t+H\\
        & \mathbf{1}^T\pi_\tau = 1 \quad \forall \tau = t+1,\dots, t+H\\
    \end{aligned}
    \label{prob:mpo_rp_convex}
\end{equation}
We solve Problem \ref{prob:mpo_rp_leastsq} with Algorithm \ref{convex_mpo_rp}.\\

\begin{algorithm}[h!]
\SetAlgoLined
\textbf{Inputs}: $\pi_{t-1}$; $\delta_\tau>0$ for $\tau=t+1,...,t+H$; $tolerance$ \;
\textbf{Initialization}: $k=0$; some $\gamma^0\in[0,1]$; $\pi_\tau^0 = \pi_{\tau-1}$ for $\tau=t+1,...,t+H$\;
 \While{(improvement on objective of Problem \ref{prob:mpo_rp_leastsq}) $ >tolerance$}{
  Solve Problem \ref{prob:mpo_rp_convex} and get optimal solution $\hat{\pi}^k = [\hat{\pi}_{t+1}^{k}, ..., \hat{\pi}_{t+H}^{k}]$\;
  $\pi_\tau^{k+1} = \pi_\tau^{k} + \gamma^k(\hat{\pi}_\tau^{k} - \pi_\tau^{k})$ for $\tau=t+1,...,t+H$ \;
  Update $\gamma^{k+1}$ \;
  $k \leftarrow k+1$ \;
 }
 \caption{Successive Convex optimization for Model Predictive Control with Risk Parity}
\label{convex_mpo_rp}
\end{algorithm}

\noindent
\cite{Feng2015} provide the convergence analysis in the case of single-period model, and the same analysis is applicable to our setting.\\
\\
In addition to multi-period formulations of mean-variance and risk parity portfolios, we implement the single period formulations to analyze the contribution of multi-period formulations. We employ the same objective functions with transaction penalties in single period formulations, but the only decision variable is $\pi_{t+1}$. Inputs are one period ahead forecasts of asset returns and covariance matrix, and the current portfolio allocation vector $\pi_t$.  

\subsubsection{Computational Performance Comparison}
The advantages of using the successive convex formulation for the multi-period risk-parity problem is two-fold: accuracy and running time. First, our experiments show that the convergence to global optimum is more stable with the successive convex programs. In particular, the algorithm usually converges within 10 steps and provides the risk-parity allocation as desired. To illustrate the accuracy of the convex algorithm on the convex formulation of the multi-period risk parity portfolio (Problem \ref{prob:mpo_rp_convex}), we run the successive algorithm in CVXPY (\cite{Diamond2016}) with OSQP solver (Operator Splitting solver for Quadratic Programs) (\cite{Stellato2020}). The original risk-parity MPC (Problem \ref{prob:mpo_rp_leastsq}) is computed with SLSQP solver (Sequential Least Squares Programming) (\cite{Kraft1988}) in SciPy package. We set maximum iteration number to 10,000 for both solvers, and left other parameters at their default values. Accuracy is measured $\ell_1$ norm of ex post and ex ante risk contributions. We consider portfolios with close to zero transaction penalty ($\gamma^{trade}\leq 10^{-3}$) to compare the the ex post risk contributions with the nominal risk-parity solution. Error metrics are reported for four planning horizons ($H=1,5,15,30$) and presented in Table \ref{tab:accuracy}. Computations are performed over the period training period (1998-2005) with 10 assets.  Our algorithm achieves a risk-parity measure of essentially zero, whereas the original formulation fails to converge from time to time. When the transaction penalty increases, our convex formulation leads to result deviating from the nominal risk-parity solution in order to balance the trade-off between risk-parity objective and turnover rates The risk contribution from each asset class stays comfortably close to each other, while providing avoidance of excessive transactions. As the planning horizon ($H$) increases, the rate of increase in errors is larger in the non-convex formulation (Problem \ref{prob:mpo_rp_leastsq}) than convex formulation (Problem \ref{prob:mpo_rp_convex}). The errors also gets larger when transaction penalty increases, that it is almost 10 times larger for convex formulation when $\gamma^{trade} = 10^{-3}$ than that when $\gamma^{trade} = 10^{-6}$. Higher transaction penalty diverges allocation from the  true risk parity portfolio. In addition, the max error of the convex formulation ($71\times10^{-4}$) is also significantly smaller than that of the non-convex formulation ($6,000 \times 10^{-4}$) over the hyperparameter space. For small $H$, the worst case error of convex formulation is one magnitude smaller than that of the non-convex formulation. For long plan-ahead horizons $H=15$ and $H=30$, the non-convex formulation fails to converge on certain days, whereas the convex formulation consistently provide the desired solution. \\

\begin{table}[hbt!]
    \centering
\begin{tabular*}{\textwidth}{c @{\extracolsep{\fill}} cccccc}
\toprule
& & \multicolumn{4}{c}{$\gamma^{trade}$}\\
\cmidrule{3-6}
{Horizon ($H$)}&{Problem Type} & $10^{-6}$ & $10^{-5}$ & $10^{-4}$ &  $10^{-3}$ \\
\midrule
&  & \multicolumn{4}{c}{\textit{Mean Error}} \\
\cmidrule{3-6}
{$H=1$} & \textit{Problem 3}  &  9.75 &  9.76 &   9.93 &  24.27 \\

&\textit{Problem 7}   &  0.22 &  0.42 &   2.88 &  24.50 \\
\midrule
{$H=5$} & \textit{Problem 3} &  8.92 &  8.95 &  11.19 &  51.28 \\
&\textit{Problem 7}   &  0.14 &  0.46 &   4.02 &  35.15 \\
\midrule
{$H=15$} & \textit{Problem 3} & 35.82 & 35.82 &  38.63 &  82.74 \\
& \textit{Problem 7}   &  0.12 &  0.44 &   4.02 &  35.48 \\
\midrule
{$H=30$} & \textit{Problem 3} & 49.20 & 49.04 &  54.16 & 102.99 \\
& \textit{Problem 7}   &  0.11 &  0.44 &   4.02 &  35.48 \\
\bottomrule
\end{tabular*}
    \caption{Errors are measured as $\ell_1$ distance of ex post risk contributions to ex ante risk contributions and reported in magnitude of $10^{-4}$. \textit{Problem 3} and \textit{7} refer to non-convex least squares and convex formulations of multi-period risk parity portfolios.}
    \label{tab:accuracy}
\end{table}

\noindent
Second, the successive algorithm successfully shorten the running time from the non-convex formulation. For a wide range of hyperparameters, the convex formulation consistently takes shorter time to converge than the original non-convex formulation. Table \ref{tab:run-time} presents the average running time per iteration for each problem. Our experiments show that for $\gamma^{trade}\geq 0.5$ the non-convex problem becomes ill-conditioned and doesn't deviate from starting allocation. Therefore, computation times are reported for the range of $\gamma^{trade}\in \{10^{-6},\dots,0.1\}$ with the same planning horizon values. Running time increases with $H$ and $\gamma^{trade}$ for both problems. Larger $H$ makes problem more complex due to longer planning horizon, and a larger transaction penalty ($\gamma^{trade}$) impacts the objective function. The difference between running times becomes more obvious for $H=15,30$ and $\gamma^{trade}\geq 0.01$. Convex formulation is 30 times faster than the non-convex formulation for $H=15, \gamma^{trade}\geq 0.05$. Furthermore, the gain in computational run time is more obvious for the longest planning horizon, $H=30$. For the largest trading penalty ($\gamma^{trade}$ = 0.1) the gain in average running speed is close to 100 times. Both problems have the highest worst case running times for $H\geq15$. However, convex formulation (10.47 sec.) has significantly smaller maximum running time per iteration than the non-convex formulation (98.75 sec.) over the training period. The convex formulation of multi-period risk parity problem improves robustness and reduces the computational time, which is helpful in hyperparameter search and backtesting processes.
\begin{table}[hbt!]
    \centering
\begin{tabular*}{\textwidth}{c @{\extracolsep{\fill}} ccccccccc}
\toprule
& & \multicolumn{7}{c}{$\gamma^{trade}$}\\
\cmidrule{3-9}
{Horizon ($H$)}&{Problem Type} & $10^{-6}$ & $10^{-5}$ & $10^{-4}$ &  $10^{-3}$ &  $0.01$ &$0.05$ & $0.1$  \\
\midrule
&  & \multicolumn{7}{c}{\textit{Mean Running Time (sec)}} \\
\cmidrule{3-9}
{$H=1$} & \textit{Problem 3} &  0.01 &  0.01 &   0.01 &  0.01 &  0.01 &  0.02 &0.03 \\

&\textit{Problem 7}  &0.02 &  0.02 &   0.02 &  0.02 &  0.02 &  0.02 & 0.02 \\
\midrule
{$H=5$} & \textit{Problem 3} &  0.12 &  0.11 &   0.11 &  0.15 &  0.19 &  0.41 & 0.55 \\
&\textit{Problem 7}  &  0.08 &  0.07 &   0.07 &  0.08 &  0.07 &  0.08 &0.08\\
\midrule

{$H=15$} & \textit{Problem 3} &  1.20 &  1.19 &   1.18 &  1.30 &  2.58 &  6.13 & 7.82 \\
& \textit{Problem 7}  &  0.21 &  0.22 &   0.22 &  0.22 &  0.24 &  0.24 &0.24\\
\midrule
{$H=30$} & \textit{Problem 3} &  5.50 &  5.64 &   5.48 &  5.41 & 11.54 & 34.47 & 47.59 \\
& \textit{Problem 7}   &  0.45 &  0.45 &   0.43 &  0.40 &  0.43 &  0.47 & 0.5\\
\bottomrule
\end{tabular*}
    \caption{Average running time per iteration reported in seconds. \textit{Problem 3} and \textit{7} refer to non-convex least squares and convex formulations of multi-period risk parity portfolios.}
    \label{tab:run-time}
\end{table}

\subsection{Investment Model Framework Specifications}
We implement the multi-period portfolio models in a rolling window whose parameters are learned based on returns of past 2000 days. Every time period, we train the HMM to estimate asset parameters and the regime transition matrix for the next period. 
In each period, we assume a transaction cost of 10 basis points for each asset class\footnote{The same transaction cost is employed in \cite{Nystrup2019}.}. We consider various investment strategies in the portfolio analysis: 1) 100\% stock, 2) equal weight ($1/n$), 3) 100\% mean-variance ETF based on single and multi periods mean-variance (SPO,MPO), and 4) 100\% risk-parity ETF based on single and multi periods risk parity (SPO\_RP,MPO\_RP). We perform the following steps for portfolio optimization techniques at each time period $t$, until the end of investment horizon $T$:
\begin{enumerate}
    \item Update HMM model parameters and get estimates of asset parameters for $H$ periods ahead
    \item Compute the optimal sequence of portfolio allocations $\pi^*_{k,t+1},\dots,\pi^*_{k,t+H}$ for each strategy $k \in$ \{SPO,MPO,SPO\_RP,MPO\_RP\} via optimization Problem \ref{mpo} and \ref{prob:mpo_rp_convex}     
    \item Execute trades for each strategy $\pi^*_{k,t+1}$ and calculate portfolio returns after transaction costs
    \item Return to step 1 and repeat
\end{enumerate}

\section{Performance on Market Data}\label{Empirical}
\subsection{Asset Classes}
\begin{figure}[hbt!]
    \centering
    \includegraphics[width=1\textwidth]{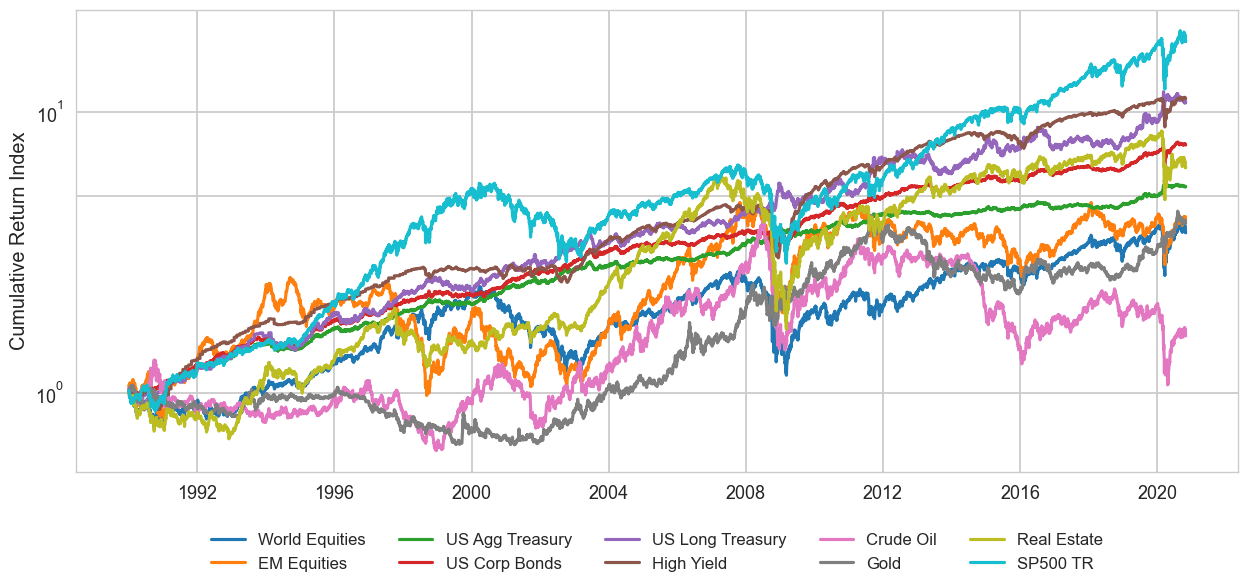}
    \caption{Asset performance over the period}
    \label{fig:asset_performance}
\end{figure}

In the portfolio analysis, we consider daily asset returns from major asset classes\footnote{World Equities: MSCI World Index (MXWO) EM Equities: MSCI Emerging Market Index (MXEF) US Domestic Equity: SP500 Total Return Index (SPXT) US Treasury: Barclays US Aggregate Treasury (LUATTRUU) US Corporate Bond : Barclays US Corporate Investment Grade (LUACTRUU) US Long Treasury: Barclays US Long Treasury (10+ years to maturity) (LUTLTRUU) US High Yield Bond: Barclays US Corporate High Yield (LF98TRUU) Crude Oil: S\&P GSCI Index (SPGSCI) Gold: LBMA Gold Price (GOLDLNPM) Real Estate: FTSE EPRA/NAREIT Developed Total Return Index (RUGL) } over the period from 1991 to 2020. 1-month Treasury bill rate returns are obtained from Kenneth R. French data library \footnote{\url{http://mba.tuck.dartmouth.edu/pages/faculty/ken.french/data_library.html}}, and used as a proxy for risk-free rate. Figure \ref{fig:asset_performance} shows cumulative asset performance, and performance statics are reported in Table \ref{asset_stat}. Notice the bond indices have the highest Sharpe ratio due to historically  macroeconomic conditions, but the US domestic equity index (S\&P 500) has the highest returns during the same period.

\begin{table}[hbt!]
\small
\centering
\begin{tabular*}{\textwidth}{c @{\extracolsep{\fill}} ccccccc}
\toprule
{} &  Ret.(\%) &  Vol.(\%) &  Max DD &  Sharpe &  Calmar &  Cum. Ret. \\
\midrule
US Domestic Equity         &          9.78 &         18.25 &          0.55 &    0.38 &          0.18 &                          1775.08 \\
World Equities   &          4.34 &         15.22 &          0.57 &    0.11 &          0.08 &                         370.43 \\
EM Equities      &          4.68 &         17.97 &          0.63 &    0.11 &          0.07 &                          410.16 \\
US Agg Treasury  &          5.62 &          4.61 &          0.07 &    0.63 &          0.78 &                           539.76 \\
US Corp Bonds    &          6.80 &          5.18 &          0.16 &    0.78 &          0.42 &                           760.91 \\
US Long Treasury &          8.01 &         10.84 &          0.18 &    0.48 &          0.43 &                          1075.64 \\
High Yield       &          8.12 &          6.01 &          0.35 &    0.89 &          0.23 &                          1111.14 \\
Crude Oil        &          1.52 &         21.73 &          0.74 &   -0.05 &          0.02 &                           159.13 \\
Gold             &          4.61 &         15.69 &          0.47 &    0.12 &          0.10 &                           401.88 \\
Real Estate      &          6.17 &         16.28 &          0.71 &    0.21 &          0.09 &                           633.56 \\

\bottomrule
\end{tabular*}
\caption{Annualized performance metrics of the ten asset indices over the period from 1990-01-02 to 2020-10-30.}
\label{asset_stat}
\end{table}

\subsection{Choice of hyperparameter}
We separate the market data so that 1991-2005 are used for hyperparameter tuning, and 2006-2020Q3 are fully out-of-sample. After taking out the first 2,000 days for HMM parameter estimation, the actual tuning period we use starts from 1998 until the end of 2005. In particular, the hyperparameters to be tuned involve:
\begin{itemize}
    \item In SPO (single-period mean-variance optimization) and MPO (multi-period mean-variance optimization) formulation, one needs to decide the mean-variance coefficient $\gamma^{risk}$ and transaction penalty $\gamma^{trade}$. In multi-period setting, the length of planning horizon $H$ should also be chosen carefully.
    \item In SPO\_RP (single-period risk-parity optimization) and MPO\_RP (multi-period risk-parity optimization) framework, one needs to choose the transaction penalty $\gamma^{trade}$, with one extra hyperparameter $H$ in the multi-period setup. This quantity could be different from that of MPO formulation, as the scales of mean-variance objective and risk-parity objective differ from each other. In the convex formulation of risk parity, the selection of hyperparameters may affect the running speed of Algorithm \ref{convex_mpo_rp}, but will not impact the output allocation from the algorithm as long as the solution converges to the proper optimum. Following \cite{Feng2015} and after some tuning, we choose $\gamma_0=0.8$ with updating rule $\gamma_k = 1-10^{-7}\gamma_{k-1}$ and $\delta_\tau=\frac{Trace(\hat{\Sigma}_{\tau|t})}{40n}$, where $n$ is the number of assets. 
\end{itemize}
Recall that the single-period model is the same as their multi-period counterpart with $H=1$. In the following subsections, we will analyze the hyperparameter tuning with mean-variance formulation and risk-parity formulation, respectively. We will choose the hyperparameters with consideration of return generation, Sharpe ratio and turnover rate, where the turnover rate measures the length on average an asset is held in a portfolio. Mathematically, the annual turnover rate is $\underset{\tau}{\Sigma}\|u_\tau\|_{1}$  where $\tau$ runs over all trading days in the year, and $u_\tau = w_\tau - w^{\rightarrow}_{\tau-1}$ is the vector of trade amount on day $\tau$.
Notice that tuning period includes a major contraction (2001-2002) period, still all strategies provide promising Sharpe ratios. Although different families of strategies can bring distinct performance under different market conditions, it is reasonable to assume that, within the same family of strategies, the relationship between hyperparameters and performance is rather stable.
\subsubsection{Hyperparameters on SPO and MPO}
\begin{figure}[hbt!]
    \centering
     \includegraphics[width = 1\textwidth]{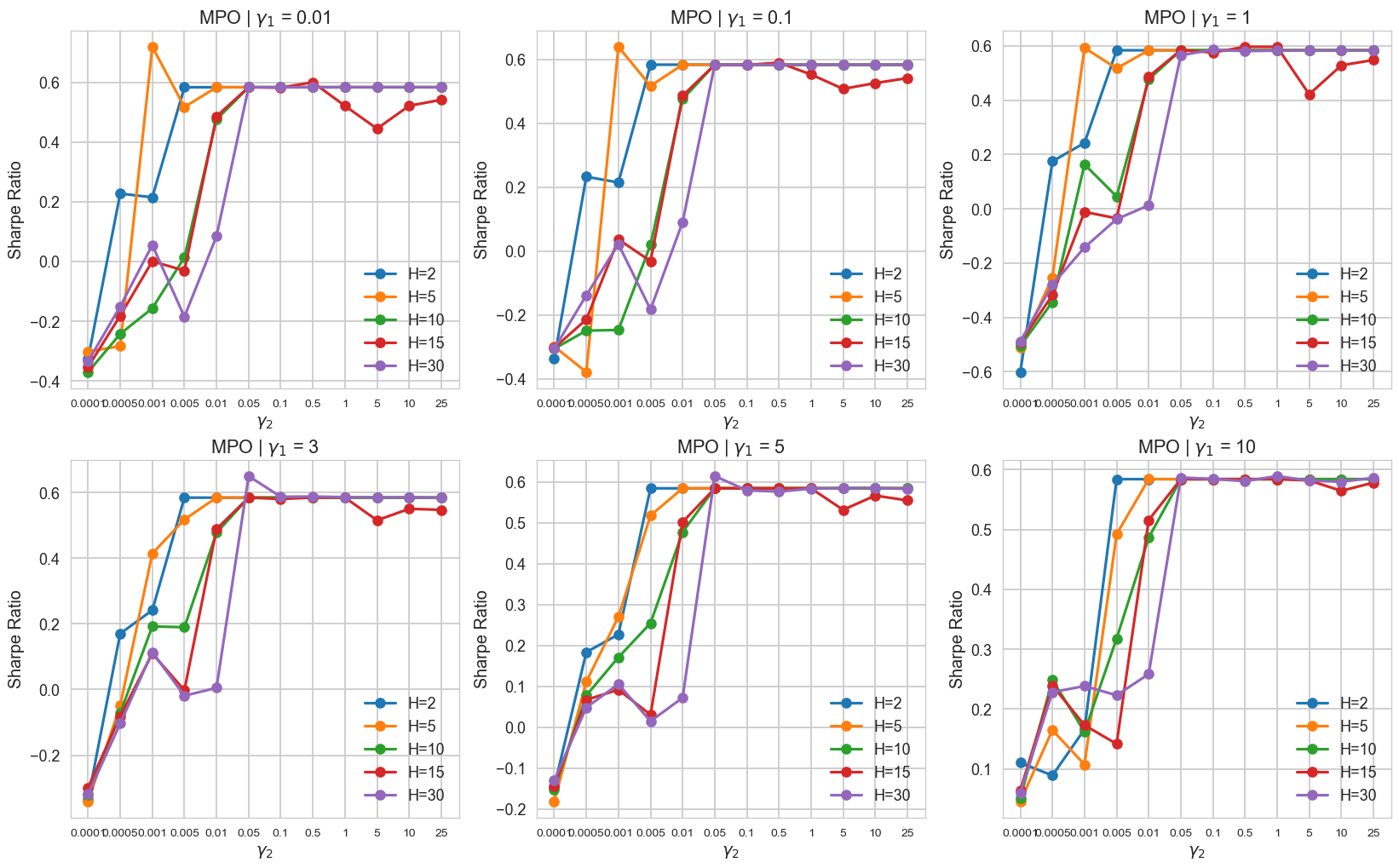}
         \caption{Hyperparameter search results for mean-variance portfolios over the period 1998-2005. $\gamma_1$ and $\gamma_2$ denote risk aversion and transaction penalty, respectively. $H$ is the planning horizon for the multi-period portfolio.}
    \label{fig:hyperparam_all_mvo}
\end{figure}

Portfolios over the tuning period with various choice of planning horizon $H\in\{1,2,5,10,15,30\}$, risk aversion coefficient $\gamma^{risk} \in \{0.01,0.1,1,3,5,10\}$ transaction penalty \lword{$\gamma^{trade} \in\{0.0001,0.0005,0.001,0.005,0.01,0.05,0.1,0.5,1,5,10,25\}$} appear in Figure \ref{fig:hyperparam_all_mvo}. First, we decide on the planning horizon of the multi-period model. We find that shorter planning horizons provide better Sharpe ratios across different transaction penalties over the training period. With MPC, since all future parameters are replaced with their forecasting values, the estimation accuracy drops as planning horizon gets longer. There is a trade-off between a longer planning horizon and better estimation of parameters. Our tuning period suggests that a planning horizon of 5 days yields good performance, and we will keep $H=5$ for $\gamma^{risk}$ and $\gamma^{trade}$ tuning as well as for the out of sample tests.\\
\\
In general, a higher transaction penalty $\gamma^{trade}$ leads to a lower turnover rate. On the other hand, we find that the turnover rate does not further decrease and Sharpe ratio stops increasing after a certain threshold, which is around 0.01 for MPO. Recall that the purpose of including the transaction penalty is to avoid frequent trading that consumes profit. Therefore, we pick $\gamma^{trade}$ at the kink, so that transactions are controlled in a meaningful way, without fully discouraging transactions and avoiding profit-generating rebalances. The choices of $\gamma^{trade}$ that are relatively close to the transaction cost of 10bp also have good interpretability with the goal of generating returns (Remark \ref{gamma_trade_remark}), though it can be higher or lower based on the investors' transaction preferences.\\

\begin{remark}\label{gamma_trade_remark}
By choosing a $\gamma^{trade}$ close to the transaction cost, the return part and transaction penalty part in the objective function of Problem \ref{mpo} approximates the actual returns. Note that the expected return in period $\tau$ with allocation $\pi_\tau$ is $\hat{r}^T_{\tau|t}\pi_\tau$, and the transaction cost paid as a proportion of total wealth is $(transaction\_cost*\|\pi_\tau - \pi_{\tau-1}^{\rightarrow}\|_1)$. Therefore, the expected realized return in period $\tau$ is $\hat{r}^T_{\tau|t}\pi_\tau - (transaction\_cost*\|\pi_\tau - \pi_{\tau-1}^{\rightarrow}\|_1)$, which is approximated by $\tau$ is $\hat{r}^T_{\tau|t}\pi_\tau - (transaction\_cost*\|\pi_\tau - \pi_{\tau-1}\|_1)$ when returns are close to zero. When $\gamma^{trade}$ is close to the linear transaction cost, the corresponding return and transaction penalty parts together approximates expected actual returns.
\end{remark}

\noindent
The $\gamma^{risk}$ parameter controls the trade-off between return and risk, and describes the subjective preference of the investor. Selecting $\gamma^{risk}$ on Sharpe ratio solely could be dangerous due to the market instability. When the market booms, small $\gamma^{risk}$ that leads to aggressive approach will generate high Sharpe ratio, whereas conservatively low $\gamma^{risk}$ usually outperforms in bearish market. In order to provide a reliable return-risk balance, we introduce an approximation of mean-variance optimization to constant relative risk aversion (CRRA) utility in Remark \ref{gamma_risk_remark}, where the CRRA utility function is $U(W)=\begin{cases}
\frac{w^\gamma}{\gamma} & \text{ if $\gamma\neq 0$ }\\
ln(w) & \text{ if $\gamma= 0$ }
\end{cases}$ at wealth level $W$ for risk aversion coefficient $\gamma$. Recalling that return estimation is generally harder and less accurate than covariance estimation, we pick a relatively conservative $\gamma^{risk}$ in order to achieve a stable portfolio.
The literature suggests that people usually have a CRRA coefficient $\gamma\in[-10,0]$, based on which we choose $\gamma^{risk}$ in Problem \ref{mpo} equal to be \textit{5}.

\begin{remark}\label{gamma_risk_remark}
With some generic assumptions, we show that there is an approximate equivalence between CRRA utility with risk aversion $\gamma$ and mean-variance optimization problem with coefficient $\lambda\approx \frac{1-\gamma}{2}$. A second order Taylor expansion gives
\begin{equation*}
    \mathbb{E}[U(W)] \approx U(\mathbb{E}[W]) + \frac{1}{2}(\gamma-1)(\mathbb{E}[W])^{\gamma-2}Var(W)
\end{equation*}
Let $m=\mathbb{E}[W]$ and $s^2=Var(W)$. Note terminal wealth $W=W_0(1+r)$ where $W_0$ is the initial wealth and $r$ is return. Thus $m= W_0(1+\Bar{r})$ and $s^2=s_r^2W_0^2$ where $\Bar{r}$ and $s_r^2$ are the mean and variance of $r$, respectively. We have
\begin{equation*}
\begin{aligned}
        \mathbb{E}[U(W)] &\approx \frac{m^\gamma}{\gamma} + \frac{1}{2}(\gamma-1)m^{\gamma-2}s^2\\
    &= W_0^\gamma \left(\frac{1}{\gamma}(1+\Bar{r})^\gamma + \frac{1}{2}(\gamma-1)(1+\Bar{r})^{\gamma-2}s_r^2\right)\\
    &\approx W_0^\gamma \left(\frac{1}{\gamma} + \Bar{r} + \frac{1}{2}(\gamma-1)\Bar{r}^2 + \frac{1}{2}(\gamma-1)s_r^2 + \frac{1}{2}(\gamma-1)s_r^2(\gamma-2)\Bar{r}+...\right)\\ 
    &\hspace{5cm} \text{ with a Taylor expansion of $\Bar{r}$ around 0}\\
    &\approx W_0^\gamma \left(\frac{1}{\gamma}+\Bar{r}+\frac{1}{2}(\gamma-1)s_r^2\right) \text{ since $\Bar{r}$ is close to 0}
\end{aligned}
\end{equation*}
Therefore, maximizing $\mathbb{E}[U(W)]$ is approximately equivalent to maximizing $\Bar{r}+\frac{1}{2}(\gamma-1)s_r^2 = \Bar{r} - \frac{1-\gamma}{2}s_r^2$.
\end{remark}

\noindent
For mean-variance optimization, we assume the initial allocation is an equal-weighted portfolio. The reason is two-fold. First, it provides a better comparison when all tested strategies start with the same given allocation. Second and more important, starting with equal-weighted portfolio generates stabler portfolio. In particular, imagine a starting allocation with zeros invested in all categories. With the constraint of all weights adding up to one, the first-period transaction penalty will always be $\gamma^{trade}*1$, and the optimal portfolio largely depends on the return and covariance estimation of the first day. On the other hand, starting with equal weights offers stability on allocation decisions and provides consistent performance over various periods. 

\begin{figure}
    \centering
    \includegraphics[width = 1\textwidth]{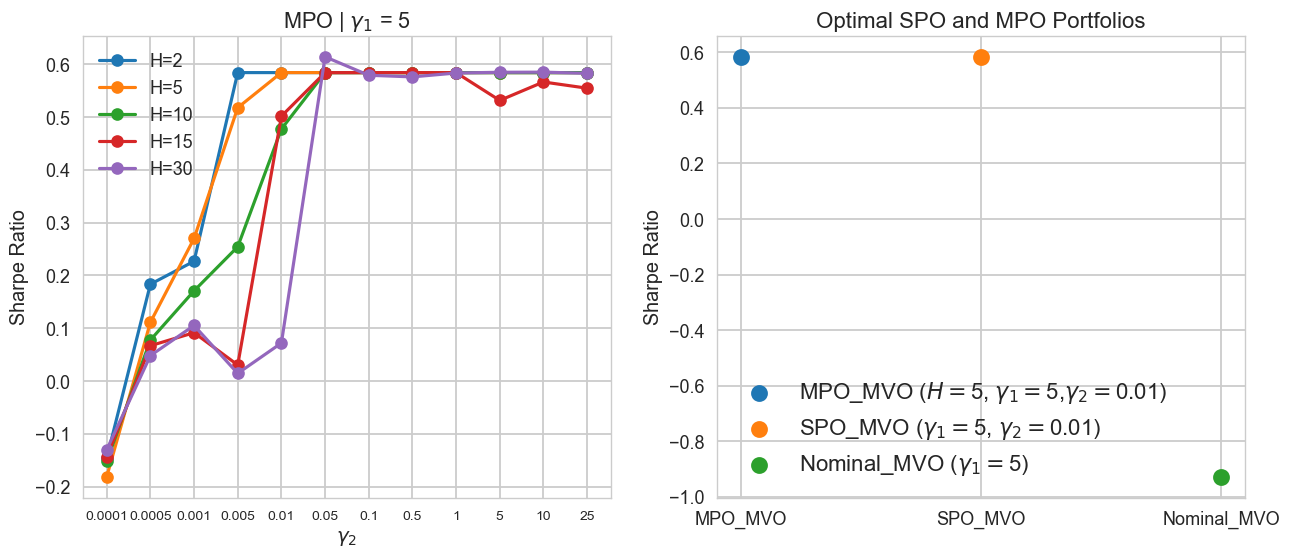}
    \caption{Optimal SPO and MPO portfolios with selected parameters. $\gamma_1$ and $\gamma_2$ denote risk aversion and transaction penalty, respectively. $H$ is the planning horizon for the multi-period portfolio. }
    \label{fig:hyperparam_all_mvo2}
\end{figure}

\subsubsection{Hyperparameters on SPO\_RP and MPO\_RP}
\begin{figure}[hbt!]
    \centering
    \includegraphics[width = 1\textwidth]{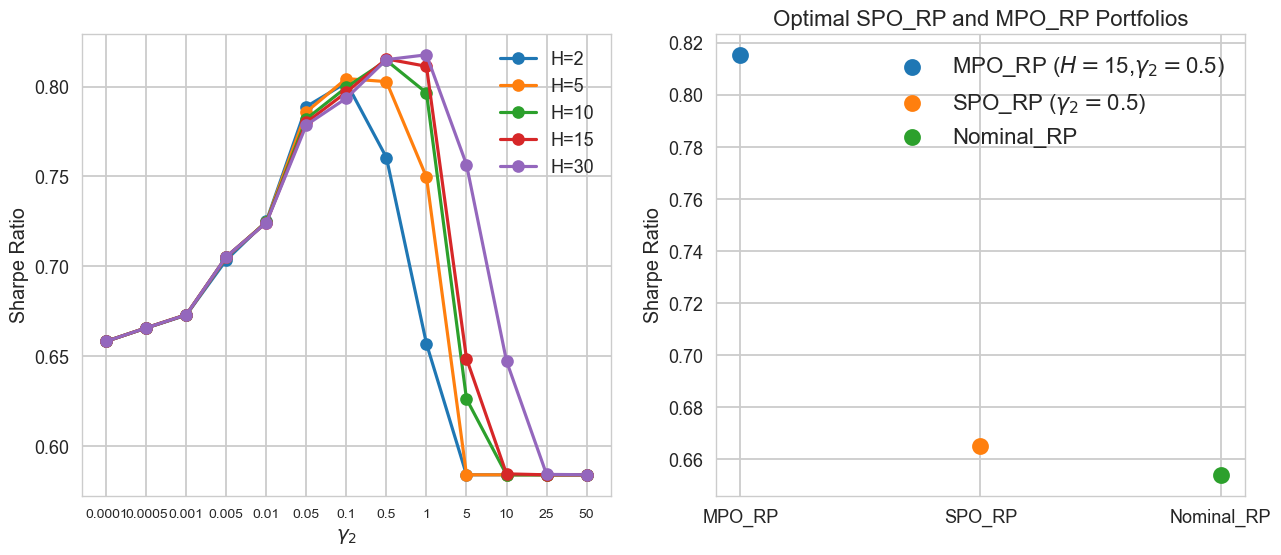}
    \caption{Hyperparameter search results for risk parity portfolios over the period 1998-2004. $\gamma_2$ denotes transaction penalty and $H$ is the planning horizon for the multi-period portfolio.}
    \label{fig:hyperparam_all_rp}
\end{figure}

The same ranges for hyperparameters $\gamma^{trade}$ and $H$ are considered in risk parity portfolios and results appear in Figure \ref{fig:hyperparam_all_rp}. When the transaction penalty is small, the penalty is negligible compared to the risk parity term, and therefore, the optimal solution satisfies nominal risk parity condition. Small transaction penalty fails to provide a meaningful control of the turnovers. 
For the multi-period risk-parity model (MPO\_RP), we choose $H=15$ and $\gamma^{trade}=0.5$. Note that in risk-parity setup, a longer planning horizon generally means a higher Sharpe ratio as our results suggest. With running time considered, we decide that $H=15$ is a reasonable choice for MPO\_RP.\\ 
\\
We observe that over the training period optimal multi-period mean-variance and risk parity portfolios have significantly better performance than their traditional single period counterparts. The transaction penalty in MPC portfolios explain the performance difference from nominal portfolios. On the other hand, we observe a slight performance gain in the multi-period portfolio over the single-period model, in Sharpe ratio and turnover rates. We believe that this is caused by model parameter estimations by the HMM model. The return estimation is a weighted average of mean return of normal and contraction regimes over the past 2000 days. When multiple asset categories are included in the model, the difference in returns under various regimes is not as obvious as the discrepancy in covariance matrices. Therefore, even though the model provides robust hints on the asset returns on a relative basis, without leveraging other micro-structure data, the return estimations based on historical returns alone are less informative as desired. 
\subsection{Comparison of performance}

Comparing the problem formulation and resulting metrics on returns with these strategies, we find that MPO has the advantage of considering return estimations, 
whereas MPO\_RP brings higher Sharpe ratio. Their relative strength is consistent with their single-period counterparties as discussed in risk parity literature (\cite{Maillard2010}). 
Both mean-variance and risk-parity objectives with transaction penalty successfully reduce the turnover rates to be below the fix-mix benchmark. Since we have chosen a relatively conservative set of hyperparameters for the mean-variance formulation, the reduction in turnover rates is significant in MPO results. In our mean-variance formulation, the transaction penalty is set equal to  $\gamma^{trade}=0.01$ to avoid unnecessary trades triggered by fake signals, which discourages any trade suggested by single-period forecast, and leads to an essentially zero turnover rates in SPO. Yet, when multi-period forecasting is considered, the trading signal is stronger than that of single-period counterpart, because there is profit by gradually switching to the desired allocation. On the other hand, a multi-period risk-parity formulation provides better transaction control than the single-period risk-parity by predicting the future covariance matrices.\\
\\
Risk parity strategy is often employed with leverage to achieve desired return level. However, leverage introduces new risks to the portfolio. Here, we consider an unlevered portfolio strategy. Both strategies provide meaningful control of volatility and turnover rates, and beat the fix-mix strategy Sharpe ratio of 0.57. MPO reduces the turnover rates by 95\% and offers a Sharpe ratio of 0.64, while MPO\_RP reduces the turnover rates by 42\% and leads to a Sharpe ratio of 0.97 over the out-of-sample period. 

\begin{table}[hbt!]
\small
    \centering
    \begin{tabular*}{\textwidth}{c @{\extracolsep{\fill}} cccccccc}
\toprule
{} &  Ann. Ret.(\%) &  Vol.(\%) &  Max DD &  Sharpe  &  Calmar  &  Turnover & Cum. Ret. \\
\midrule
S\&P 500      &         9.67 &        20.31 &         0.55 &   0.42 &         0.18 &          0.00 &            395.78 \\
Fix-mix (1/n)               &        6.11 &        8.69 &        0.35 &  0.57 &        0.17 &         1.43 &           241.81 \\
SPO                   &        5.71 &        8.04 &        0.34 &  0.57 &        0.17 &         0.00$^{\mathrm{a}}$ &           228.54 \\
MPO                   &        5.28 &        6.42 &        0.31 &  0.64 &        0.17 &         0.07 &           215.26 \\
SPO\_RP                &        6.18 &        8.06 &        0.33 &  0.62 &        0.19 &         0.05 &           244.15 \\
MPO\_RP                &        6.02 &        5.00 &        0.22 &  0.97 &        0.27 &         0.83 &           238.78 \\
\bottomrule
\multicolumn{8}{p{\textwidth}}{$^{\mathrm{a}}$ In the mean-variance formulation, our choice of transaction penalty is conservative with $\gamma^{trade}=0.01$. When single-period forecasting is considered, the signal is not strong enough to trigger a trade, leading to a low turnover rates in SPO. On the contrary, when multi-period forecasting is considered, there is advantage of gradually switching to the desired allocation, resulting in a non-zero turnover rate in MPO.} \\
\end{tabular*}
\caption{Annualized portfolio metrics over the period from 2006-01-01 to 2020-11-30 }
\label{tab:stats_mpo_spo}
\end{table}
 In addition, we find that the resulting allocations from the MPO and MPO\_RP strategies are different, and have relative strength and weakness under various market conditions, which is consistent with their objectives by nature. We will zoom in a recent period and compare the allocations in detail in Section \ref{covid_performance}. 
 
 \subsubsection{Model Performance without HMM}
 To justify the use of HMM, we compare the performance of the same set of strategies with parameters estimated with and without HMM. Model performances without HMM inputs are presented in Table \ref{tab:withouthmm}. We replace the forecast of expected return and covariances by their historical average in all the models considered, with the same set of hyperparameters in the previous subsection. In this setting, no regime information is involved, and the forecasting values are the same for all looking-forward periods. It turns out that when parameter estimation for all future periods are the same, SPO and MPO provides similar portfolios when the hyperparameters $\gamma^{risk}$ and $\gamma^{trade}$ are relatively conservative as we have chosen. When the returns are estimated with average past returns without utilizing regime concept in HMM, the estimation of returns are close to each other. With a conservatively chosen $\gamma^{trade}=0.01$, there is barely any signal strong enough to trigger a trade. Therefore, without utilizing HMM, both the single- and multi-period mean-variance formulations mimic the buy-and-hold strategy with initial allocation equalling the 1/n portfilio. MPO\_RP, on the other hand, still beats SPO\_RP by considering ahead. In general, all models that we consider in the paper generates higher Sharpe ratio when using the parameters are estimated with HMM. 
 \begin{table}[hbt!]
\small
    \centering
\begin{tabular*}{\textwidth}{c @{\extracolsep{\fill}} cccccccc}
\toprule
{} &  Ret.(\%) &  Vol.(\%) &  Max DD &  Sharpe &  Calmar &  Turnover  &Cum. Ret. \\
\midrule
S\&P 500     &          9.67 &         20.31 &          0.55 &    0.42 &          0.18 &           0.00 &             395.78 \\
Fix-mix (1/n)      &          6.11 &          8.69 &          0.35 &    0.57 &          0.17 &           1.43 &             241.81 \\

SPO           &          5.76 &          8.07 &          0.34 &    0.57 &          0.17 &           0.00$^{\mathrm{a}}$ &            230.17 \\
MPO         &          5.76 &          8.07 &          0.34 &    0.57 &          0.17 &           0.00$^{\mathrm{a}}$ &            230.17 \\

SPO\_RP  &         6.01 &          8.19 &          0.34 &    0.59 &          0.18 &           0.01 &             238.55 \\
MPO\_RP &         6.07 &          5.17 &          0.24 &    0.95 &          0.25 &           0.19 &             240.42 \\

\bottomrule
\multicolumn{8}{p{\textwidth}}{$^{\mathrm{a}}$ The return estimation without HMM barely provide a signal strong enough to trigger a trade. Therefore, SPO and MPO essentially mimic the performance of buy-and-hold strategy with starting point equalling the 1/n portfolio.} \\
\end{tabular*}
\caption{Annualized portfolio metrics over the period from 2006-01-01 to 2020-11-30 without HMM inputs}
\label{tab:withouthmm}
\end{table}

\subsubsection{Performance during the COVID-19 Crisis}\label{covid_performance}
\begin{figure}[hbt!]
    \centering
    \includegraphics[width = 1\textwidth]{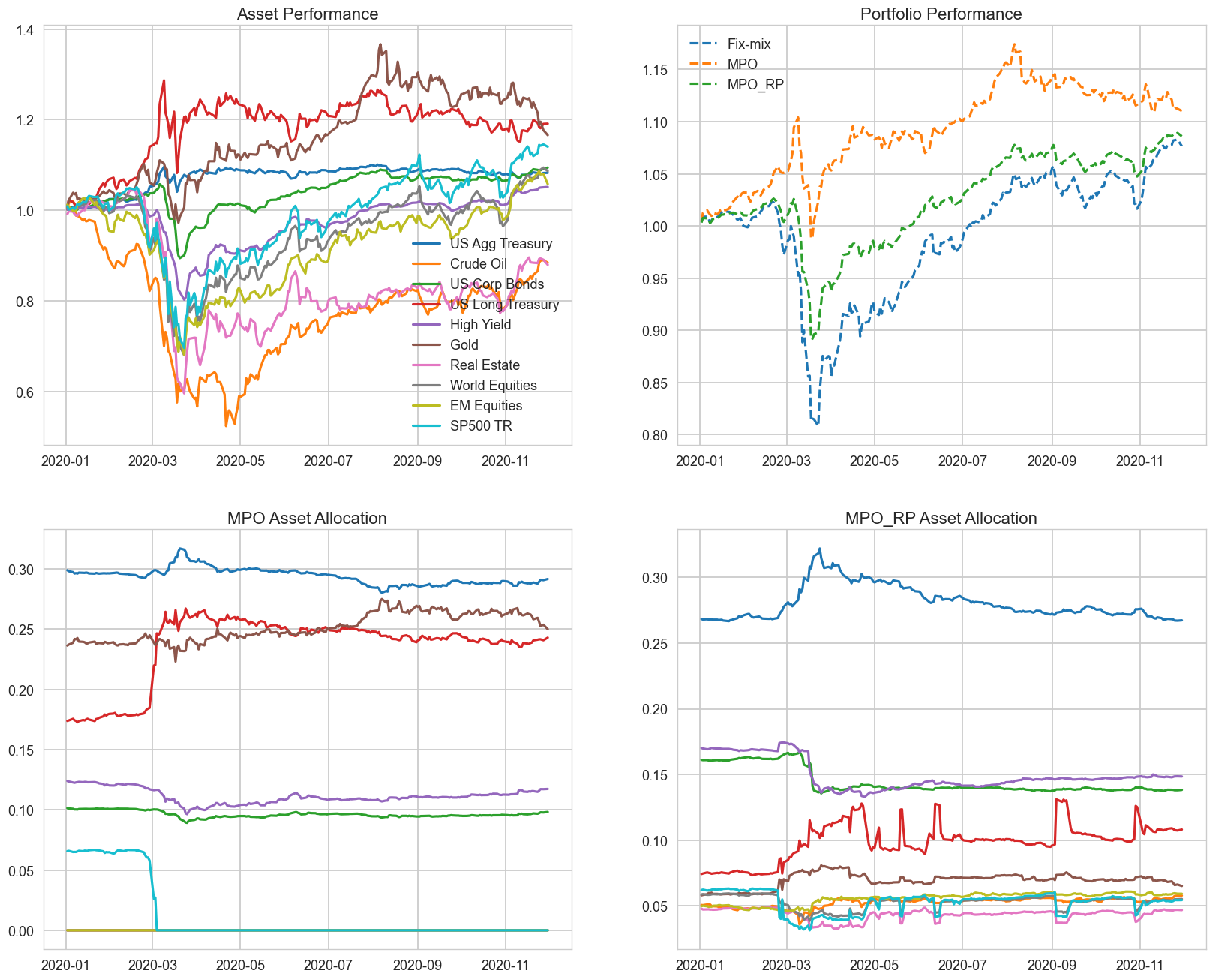}
    \caption{MPO and MPO\_RP ETF performances and asset allocations during the COVID-19 market downturn. Both MPO and MPO\_RP are successful in picking the top five performers, and outperform fix-mix strategy. MPO invests heavily in the asset classes with high returns, and MPO\_RP invests in those with low volatilities. In particular, MPO allocates more in US Long Treasury than MPO\_RP does, because it is a high-return high-volatility asset class, leading to an outperformance of MPO in this particular period.}
    \label{fig:covid performance}
\end{figure}

To understand the relative strength of the proposed strategies, we zoom in the recent period to compare the allocation and performance. Figure \ref{fig:covid performance} presents portfolio performances and asset allocations during the market downturn in March 2020 due to COVID-19 pandemic. The first plot shows that bond indices have the best performance, and long maturity government bonds have the highest cumulative return. Both MPO and MPO\_RP are successful in investing the top performers in this period, and they both allocate heavily in US Aggregate Treasury which turns out to be the most stable winner. \\
\\
Despite the fact that both strategies selects the asset classes which performed well in the COVID-19 crash, their allocations are different given their objectives and formulation. By taking return forecasting into consideration, MPO weighs heavily in the top three asset classes in terms of return, and is also able to identify the second-tier in which it invests lightly. None is allocated to the four asset classes that performed worst. On the other hand, MPO\_RP picks the asset classes that are stable while providing relatively high returns. In particular, MPO\_RP invests less in the best performer, US Long Treasury, than MPO, due to the high volatility of nature of this asset class (Table \ref{asset_stat}). The losers crashed heavily, leading to high short-term volatility, which helps risk-parity strategy to not invest heavily in them. \\
\\
Comparing the overall performance of MPO and MPO\_RP during this period, we find that both outperforms the fix-mix strategy by identifying the winners. The MPO strategy bests MPO\_RP up to mid-late March when the market downturn happened, thanks to its vast investment in US Long Treasury, an asset class with high-return and mid-high volatility during contraction periods.\\
\\
Over this sample period, both MPO and MPO\_RP allocate capitals in consistency with their objective functions. The mean-variance optimization finds a balance between chasing returns and controlling risk; whereas the risk-parity objective prefers the stably growing asset classes.

 \begin{table}[hbt!]
\small
    \centering
\begin{tabular*}{\textwidth}{c @{\extracolsep{\fill}} cccccccc}
\toprule
{} &  Ret.(\%) &  Vol.(\%) &  Max DD &  Sharpe &  Calmar &  Turnover  &Cum. Ret. \\
\midrule
S\&P 500      &        15.38 &        35.87 &         0.34 &   0.42 &         0.46 &          0.00 &            114.02 \\
Fix-mix (1/n)      &         8.26 &       15.01 &        0.21 &  0.52 &        0.39 &         2.06 &           107.54\\

SPO          &         7.85 &       13.64 &        0.19 &  0.54 &        0.41 &         0.03 &           107.18 \\
MPO         &        12.13 &        9.57 &        0.11 &  1.22 &        1.14 &         0.13 &           111.07 \\

SPO\_RP &         8.84 &       13.19 &        0.19 &  0.64 &        0.47 &         0.13 &           108.07 \\
MPO\_RP &         9.31 &        8.65 &        0.13 &  1.02 &        0.71 &         1.29 &           108.50 \\
\bottomrule

\end{tabular*}
\caption{Annualized portfolio metrics over the period from 2020-01-01 to 2020-11-30}
\label{tab:covid performance}
\end{table}

\subsection{Sensitivity to Transaction Costs}
\begin{figure}[hbt!]
    \centering
    \includegraphics[width = 1\textwidth]{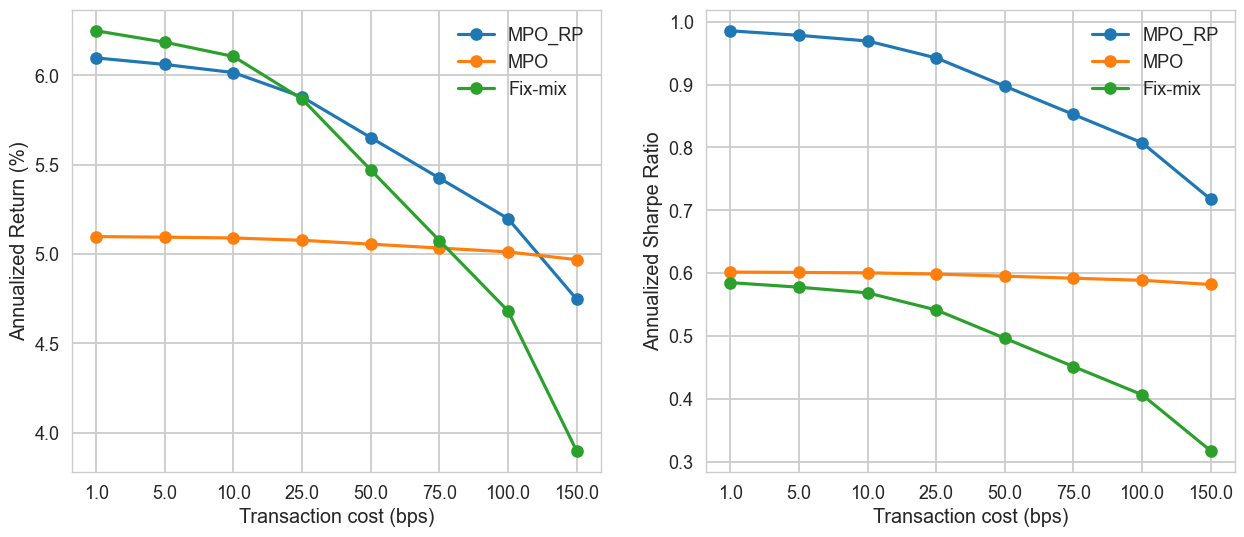}
    \caption{Performance of MPO and MPO\_RP under different transaction costs over the test period (2006-2020)}
    \label{fig:tcost-sensitivity}
\end{figure}

One natural question in evaluating allocation strategy performance is how sensitive the performance is to transaction costs. An underestimation of transaction cost may lead to a strategy that seems to be profitable, but fails to generate returns in real-world applications. To test the robustness of our proposed frameworks, we implement the same allocation strategy with a various transaction costs, ranging from 1 bp to 150 bps (Figure \ref{fig:tcost-sensitivity}). The risk-parity formulation (MPO\_RP) consistently provide higher Sharpe ratio than the mean-variance formulation (MPO) and the fix-mix benchmark. On the other hand, MPO is hurt less by unexpectedly high transaction costs due to its low turnover rates.\\
\\
For investors facing a higher range of transaction costs, we suggest re-tune the hyperparameters of both MPO and MPO\_RP to obtain a suitable choice for the transaction costs. The performance of both frameworks are robust when the estimation error of transaction costs is within a reasonable range.
\newpage
\section{Conclusion}\label{conclusion}
In this paper, we propose a risk-parity portfolio that can be effectively solved by model predictive control, and provide a successive convex program algorithm that provides faster and more robust solutions.  The risk-parity MPC generates higher Sharpe ratio than the MPC with mean-variance objective, with both formulations enjoys different strength. In the out-of-sample period, MPO and MPO\_RP leads to Sharpe ratio of 0.64 and 0.97 after transaction costs, respectively, outperforming the fix-mix benchmark of 0.57. Zooming in the recent market downturn due to COVID-19 in 2020, we observe different strength of mean-variance and risk-parity objectives. In addition, the proposed strategies are robust even when there is misspecification of transaction costs, due to their successful control of unnecessary transactions. 
\subsection{Future Work}
There are numerous next steps that we want to point out for future research. In this paper, we choose the hidden Markov model to describe asset dynamics, but other approaches can be employ for asset parameter estimations such as factor models. A good predictive model for asset parameters will improve strategy performances, especially for multi-period mean-variance portfolio. Leverage and other budget constraints can be introduced to enhance performances in the mean-variance strategy. However, budget constraints will create computationally difficulty in risk parity portfolio. In addition, we believe that a good predictive signal of cross-asset  momentum will provide substantial gains in ETF switching strategies.

\clearpage
\newpage
\bibliographystyle{apalike}
\bibliography{ref}
\newpage

\end{document}